\documentstyle[aps,prd,preprint]{revtex}
\tightenlines
\newcommand{\beq}{\begin{equation}}
\newcommand{\eeq}{\end{equation}}

\def\lap{\lower.5ex\hbox{$\; \buildrel < \over \sim \;$}}
\def\gap{\lower.5ex\hbox{$\; \buildrel > \over \sim \;$}}

\input epsf
\begin{document}

\title{Formation and Evolution of Cosmic D-strings}

\author{Gia Dvali$^1$ and Alexander Vilenkin$^2$}

\address{
$^1$ Center for Cosmology and Particle Physics, Department of Physics, New York 
University, New York, NY 10003\\
$^2$ Institute of Cosmology, Department of Physics and Astronomy,\\
Tufts University, Medford, MA 02155, USA}

\maketitle

\begin{abstract}

We study the formation of $D$ and $F$-cosmic strings in $D$-brane
annihilation after brane inflation.  We show that $D$-string formation
by quantum de Sitter fluctuations is severely suppressed, due to
suppression of RR field fluctuations in compact dimensions.  We
discuss the resonant mechanism of production of $D$ and $F$-strings,
which are formed as magnetic and electric flux tubes of the two
orthogonal gauge fields living on the world-volume of the unstable
brane.  We outline the subsequent cosmological evolution of the $D-F$
string network.  We also compare the nature of these strings with the
ordinary cosmic strings and point out some differences and
similarities.

\end{abstract}

\section{Introduction}

In the brane inflation scenario \cite{DT,Dinf}, the inflationary expansion
is driven by the attractive interaction potential between a stack of
parallel $D$-branes and a stack of anti-$D$-branes evolving in a
higher-dimensional bulk space.  The two stacks are slowly pulled
towards one another, until they collide and partially annihilate.
Brane collision marks the end of inflation. We live on one of the
surviving branes.  Annihilation proceeds via tachyon condensation
\cite{sen}.  During inflation, the tachyon is trapped in the false
vacuum, and gets destabilized only after the branes approach one
another at the string scale distance. So, in this respect
brane-anti-brane inflation naturally realizes the idea of hybrid
inflation\cite{hybrid}, in which the role of the second field is
played by the tachyon.

Brane annihilation can result in the formation of lower-dimensional
$D$-branes, which are seen as domain walls, strings, or monopoles by
brane observers.  This was discussed by Sarangi and Tye \cite{ST} (ST)
(see also \cite{JST}), who concluded that cosmic strings are copiously
produced in this scenario, while the formation of domain walls and
monopoles is strongly suppresed.  On the other hand, Majumdar and
Davis \cite{MD} (MD) have argued that all kinds of branes will be
produced in large numbers, so the usual problems associated with
cosmic overproduction of domain walls and monopoles may arise.

In the present paper we reexamine the problem of topological defect
(such as cosmic string) production in brane annihilation, with
conclusions somewhat different from those in ST and MD. The role of
the defects can be played by lower-dimensional $D$-branes, as well as
by the fundamental ($F$) strings, and we shall consider both cases.
The possibility of cosmic $F$-strings was suggested by Witten
\cite{wittencosmic}.  We discuss the production mechanisms and the
cosmological evolution of $D-F$-string networks in the context of
$D$-brane driven inflation and brane annihilation.

We begin in the next Section with a brief review of brane inflation.
Different mechanisms of defect formation at the end of brane inflation
are discussed in Sections III-VI.  
In Section VII we discuss some unusual features of $D$-strings and show
that they share some properties of both global and gauge strings. The
evolution of interacting $F$- and $D$-string networks is discussed
in Section VIII.  Our conclusions are summarized and discussed in
Section IX.

During the completion of this work, we became aware of an independent 
work by Copeland, Myers and Polchinski\cite{joe}, which also deals with 
cosmological implications of $D$ and $F$-strings.

\section{Brane inflation}

We shall assume that all extra spatial dimensions are compactified.
The parent branes fill the three large dimensions and may wrap around
some of the compact dimensions.  The two stacks of branes are
separated in the remaining $d_\perp$ dimensions, which are
orthogonal to the branes.  The role of the inflaton in this
scenario is played by the inter-brane separation.  In generic models, the size
of transverse extra dimensions $l_\perp$, the string scale $M_s$, and
the Hubble expansion rate $H$ satisfy 
\beq
H\ll l^{-1}_\perp\ll M_s.
\label{scales}
\eeq

The key idea of brane inflation is that the flatness of the inflaton
potential may be protected by the {\it locality} in the extra
space. That is, as seen from high-dimensions, the inflaton potential
is simply an inter-brane potential. The latter is generated from the
bulk (closed string) exchanges, which die-off as an inverse power of
the distance $r$, in accordance with the high-dimensional Gauss
law.  Note that for a large separation $r \gg 1/M_s$, only
massless modes (graviton, dilaton, and Ramond-Ramond fields)
contribute.  Hence, generically the inter-brane potential is
\beq
V(r)=M_s^4[A-B(M_s r)^{-(n_\perp -2)}],
\label{potential}
\eeq
where the coefficients $A$, $B$ can
be expressed in terms of the sizes of the longitudinal and transverse
compact dimensions in units of $M_s^{-1}$.
Their precise values depend on how much the system departs from
the PBS limit. For instance, with only parallel $D$-branes and 
no anti-branes, the potential vanishes, since
graviton-dilaton attraction is exactly compensated by RR-repulsion. 
This will change if one departs from the supersymmetry limit, which can be
achieved, e.g., by rotating branes at angles, or by some other
SUSY-breaking dynamics. In the present work, we shall only be interested
in the case of maximal departure from the BPS limit, due to the presence 
of parallel brane-anti-brane pairs
(their numbers can be different though).

A possible string theory implementation of inflation driven by a
system of branes and anti-branes was studied in the interesting recent
work \cite{strings,string1}.  Especially nontrivial is the issue of
consistent stabilization of moduli during inflation. In the present
work, we will only be interested in the final stage of brane
collision, and will simply assume that it was preceeded by a stage of
sufficiently long inflation, about which we shall only assume some
very general constraints, such as, e.g., Eq.~(\ref{scales}).

\section{$D$-defect formation by quantum fluctuations}

Consider a stack of branes and a stack of anti-branes separated by a
distance $r$ and moving towards one another.  When $r$ becomes
comparable to the string scale, $r\sim M_s^{-1}$, a tachyonic
instability develops. Tachyon is the state of an open string that
connects brane and anti-brane.  It has been conjectured that the brane
annihilation can be described in terms of condensation of this
tachyon\cite{sen}.  The tachyon develops a vacuum expectation value
$\chi$, which takes values in the vacuum manifold ${\cal M}$.  The
homotopy of the vacuum manifold can be conveniently described by 
$K$-theory\cite{witten}.  For example, for $N$ branes and $N$
anti-branes, ${\cal M}= U(N)$.  The corresponding homotopy groups
are\footnote{All other homotopy groups are trivial.}
\beq
\pi_{2k-1}(U(N))=Z 
\label{pi}
\eeq for $N\geq k$, indicating that ${\cal M}$ is topologically
nontrivial. Thus, there must exist topological knots. These knots are
the lower-dimension $D$-branes.  From the point of view of the low
energy four-dimensional theory, they can look as topological
defects of different co-dimension, such as walls, strings, or
pointlike charges.  (It was suggested recently\cite{dtstrings} that
from the point of view of the low-energy supergravity theory,
$D$-strings may be understood as BPS $D$-term strings, also indicating
that the energy of the parent brane-anti-brane system is of the
$D$-term type.)

 Thus, in other words, daughter branes are formed as topological defects as a
result of symmetry breaking in brane annihilation.  The tachyon
transforms as a bi-fundamental representation $(N, \bar N)$ under the
original symmetry group $U(N) \times U(N)$. This group is broken to a
diagonal subgroup $U(N)$ by the tachyon condensation.  Topologically
non-trivial knots are the vortices formed by the tachyon.  Take, for
instance, a single pair of parent brane-anti-brane.  The symmetry
group is $U(1)_1\times U(1)_2$ and the tachyon is charged under the
linear combination $A \, = A^{(1)} \, -
\, A^{(2)}$, which gets Higgsed by the condensate.  The elementary
vortices are topologically stable configurations around which the
phase of the tachyon VEV changes by $2\pi$.  They carry one unit of
$A$-magnetic flux, and are therefore charged under RR fields. Their
charge can be inferred from the WZ couplings between close and open
string modes on the world-volume of the parent branes,
\begin{equation}
\label{charge}
 \int_{p\, + \,1} \,  F\wedge \, C_{p\, - \, 1},
\end{equation}
where $F$ is the gauge field strength.  This coupling accounts for the correct  
RR charge of the
vortex under the $C_{p\, - \, 1}$-form, equal to the RR charge of the $D_{p\, 
-\, 2}$-brane.

Eq.(\ref{charge}) carries important model-independent information
about the nature of daughter defects. It is immediately clear that
daughter branes cannot be pointlike as viewed from the three
non-compact dimensions.  To see this, suppose the initial parent
branes fill $p\, = \, 3\, + \, n$ space dimensions, $n$ of which
obviously have to be compact.  The daughter branes are then charged
under the $n\, +\,2$ RR form. From the perspective of the 3
non-compact dimensions, the daughter branes will look as stable
point-like defects ($D_0$-branes) only if $C_{n\, + \, 2}$-form has a
single index in our $3+1$-dimensions.  But this is impossible, since
the number of the remaining $n\, + \, 1$ indices exceeds the number of
the compact world-volume dimensions of the parent branes.  Hence,
irrespective of the cosmological context, $0$-branes are never formed
in D-brane annihilation after inflation.

Interestingly, formation of objects looking like strings in
$3+1$-dimensions (effective $D_1$-branes) is not forbidden. Indeed,
strings are charged under the two-form fields, meaning that $C_{n\, +
\,2}$ form should have two indices in our dimensions. The remaining $n$
indices are just enough to fill the compact world-volume dimensions of
the parent branes.

Unfortunately, the same reasoning also permits the formation of 
domain wall type defects (i.e., $D_2$-branes).  Indeed, $2$-branes
will be stable if they couple to a $3$-form RR field. Hence, $C_{n\, +
\, 2}$ should have $3$ indices in our dimensions and the rest in the
compact ones. This is certainly possible,
and implies that the gauge field strength $F$ should have one index in
the extra space and one in ours.

Hence, the formation of walls versus strings becomes a dynamical
question, which we would like to address in the cosmological context.
To make the problem more transparent, we shall first reduce it to bare
essentials.  For definiteness, let us do this explicitly for the $n=1$
case, as the generalization to higher $n$ is trivial. Let $A$ be a
world-volume gauge field Higgsed by the tachyon VEV. And let $\theta$
be the phase of the tachyon $\chi$.  We shall expand $A$, $\theta$ and
the $C_{3}$ form in the Kaluza-Klein (KK) states with respect to the
compact world volume coordinate, which we shall call $x_5$, and write
down the effective $3 + 1$-dimensional couplings. The terms of our
interest are
\begin{eqnarray}
\label{kkform}
m^2\,(A_5^{(m)} \, - \, \theta^{(m)})^2 |\chi|^2 \, + \,
(A_{\mu}^{(m)}\, - \, \partial_{\mu} \theta^{(m)})^2 |\chi|^2\, + \,
m^2 (A_{\mu}^{(m)} \, -\, \partial_{\mu}\, A_5^{(m)})^2 \, +
\nonumber\\
\, m\, (A_{\mu}^{(m)} \, -\, \partial_{\mu}\, A_5^{(m)})\, 
C_{\nu\alpha\beta}^{(m)}\epsilon^{\mu\nu\alpha\beta}\, 
+\, (\partial_{\mu} A_{\mu}^{(m)} \, -\, \partial_{\nu}\,
A_{\mu}^{(m)})\, C_{5\alpha\beta}^{(m)}\epsilon^{\mu\nu\alpha\beta}.
\end{eqnarray} 

Here, $m$ is
quantized in units of $1/l_{\parallel}$, where $l_\parallel$ is the
size of the compact $x_5$-dimension.  The usual kinetic terms of
$A_{\mu}$ and $C$ are not displayed, and we have suppressed the 
relative powers of the string scale infront of the last two terms. 

 Notice that in the above
notations $A_5^{(m)}$ and $\theta^{m}$ are dimensionless, so that the
canonically normalized fields are $mA_{5}^{(m)}$, and
$\theta^{(m)}|\chi|$ respectively.  The meaning of each of the terms
in (\ref{kkform}) is clear.  What we are dealing with is the usual
Higgs effect. At each mass level $A_{\mu}^{(m)}$ eats up a combination
of $A_5^{(m)}$ and the tachyon phase $\theta^{(m)}$, and becomes a
massive spin-1 field of mass $m^2 \, + \, |\chi|^2$. Another
combination of $A_5^{(m)}$ and $\theta^{(m)}$ becomes a massive scalar
of mass $|\chi|^2$.  The important couplings are the last two terms in
(\ref{kkform}), which tell us how the objects charged under $C$ can be
created.

Let us first consider the first of these terms, which controls the
formation of 2-branes.  It is clear that in order to create a source
for the $C_{\nu\alpha\beta}$ 3-form, we either have to excite
$A_{\mu}^{(m)}$ (that is, create gradients of the gauge field in extra
dimensions), or to produce gradients of $A_{5}^{(m)}$ in our
non-compact dimensions.  For $m\neq 0$ such possibilities are costly
in energy and cannot be achieved by de Sitter fluctuations, since $H
\ll 1/l_{\parallel}$.  So the only option would be to create gradients
in the zero mode $A_{5}^{(0)}$. At the tree level this is a massless
field, and such gradients are possible, but they cannot lead to stable
defects, as long as $A_{5}^{(0)}$ stays massless.  For $2$-branes to
become stable, $A_{5}^{(0)}$ should get a potential with
discretely-degenerate minima. For instance, a periodic potential of
the form cos$(A_{5}^{(0)})$ would lead to the formation of stable
$2$-branes similar to axionic-type domain walls.  $A_{5}^{(0)}$ will
change by $2\pi$ through such a wall, and so will the field strength
of $C$, by the amount proportional to the change in $A_5^{(0)}$
\cite{DvVi}. Such potentials may indeed be generated from some
non-perturbative dynamics, but this is highly model-dependent.

Let us now turn to the last term in Eq.(\ref{kkform}) which is
responsible for the existence of $D$-strings.  We see that strings can
be formed without exciting any of the KK states of either the gauge
field or the tachyon phase. 
%Hence, no fluctuations in the compact
%world volume dimensions are required. 
We could simply direct the
magnetic flux of the zero mode gauge field $F_{\mu\nu}$ into the
non-compact directions.  As it is clear from (\ref{kkform}), this flux
would carry an effective two-form $C_{5\alpha\beta}$ RR charge, and be
stable. Hence, we might conclude that formation of cosmic strings is
not costly in energy and can be afforded by quantum fluctuations
during brane inflation.  This would be a false impression, however, since we
would be forgetting the bulk nature of the RR fields.  There are
additional $6-n$ dimensions orthogonal to the brane, in which the
$C_{n\, + \, 2}$ RR field propagates.  Directing the $F_{\mu\nu}$ flux
in three non-compact dimensions necessarily implies the existence of
RR gradients in transverse dimensions.  In other words, there is a
"zero sum" game: we can either avoid gradients of the world-volume
fields (such as $A$ and $\theta$) in $n$ world-volume directions
tangent to the parent branes, or the gradients of the RR fields in
$6-n$ transverse dimensions, but we cannot avoid both!  As a result,
the formation of cosmic $D$-strings by de Sitter fluctuations is
strongly suppressed.

\section{The role of the RR fields}

We wish to study the formation and the physical nature of $D$ and $F$ strings
in the cosmological context.  
%The argument that is usually
%given for the formation of daughter branes in brane annihilation is as
%follows.  
If the values of $\chi$ are uncorrelated beyond some correlation
length $\xi$, daughter branes can be formed as topological defects by
the usual Kibble mechanism \cite{Kibble}.  The spatial variation of
the tachyon may either be due to quantum de Sitter fluctuations during
inflation, or to thermal or resonant excitation during brane
annihilation.  We shall first consider the possibility that quantum
fluctuations are the dominant effect, as it was assumed in ST and MD.
During the course of inflation, scalar fields of mass $m \lesssim H$
fluctuate by $\delta\phi\sim H/2\pi$ on the horizon scale $H^{-1}$ per
Hubble time $H^{-1}$.  The characteristic wavelength of the
fluctuations is thus $\sim H^{-1}$.

The initial setup considered by MD \cite{MD} is that of space-filling
$D9$ and anti-$D9$-branes, and the effects of compactification were
ignored.  By the standard Kibble argument it was concluded that all
defects that can be formed will be formed as a result of parent brane
disintegration, with a typical defect separation $\sim H^{-1}$.
Note that it follows from Eq.(\ref{pi}) that the codimension of daughter
branes with respect to the parent branes must be an even integer.
MD comment that similar conclusions apply to the
case of annihilating branes in the brane inflation scenario, but this
case is not discussed in detail.

It was, however, noted by ST \cite{ST} that the effect of
compactification on brane annihilation can be very
significant. According to Eq.(\ref{scales}), the compactification
radius is small compared to the horizon, $l_\perp \ll
H^{-1}$, suggesting that the variation of $\chi$ in the compact
dimensions should be suppressed.  If $\chi$ varies only along the three
non-compact dimensions of the parent branes and not along their compact
dimensions, then the resulting daughter branes will wrap around the
same compact dimensions as the parent branes.  The codimensions of the
daughter branes should then lie within the large dimensions, and since 
the codimension should be even, it follows that the defects should be
of codimension 2, that is, cosmic strings \cite{ST}.
%\top

We believe, however, that this analysis does not fully account for the
effects of compactification, particularly in the directions transverse
to the parent branes.  $D$-branes are bulk animals, and to analyze
their formation it is not enough to consider only the dimensions
within the parent branes. As discussed above, the crucial point is
that the stable daughter D-branes carry conserved charges under the RR
fields. These come from the closed string sector and their flux
necessarily extends to all compactified directions, irrespective of
the plane of the daughter brane localization.  If the dominant source
of fluctuations are de Sitter fluctuations, then daughter branes
cannot be formed, unless these fluctuations prepare suitable gradients
of the RR fields. As we shall see, for the values of the Hubble
parameter compatible with $D$-brane-inflation, creation of such
gradients is costly in energy, so that daughter $D$-strings formation
is severely suppressed.

To be maximally general, let us consider the
formation of $D_{p}$-branes, in annihilation of a $D_{p\,+\,n}\, - \,
\bar{D}_{p\, +\, n}$ pair.  $D_{p}$ branes are charged under the $p\,
+ \, 1$ form RR field $C_{p\, + \, 1}$.  
Since $D$-brane inflation happens due to relative motion of branes,
the initial branes cannot wrap all the compactified dimensions.  
%As discussed above, 
The same is true about the daughter branes which are forced to lie
entirely within the parent brane world-volume.  Thus, in order for the
daughter $D_{p}$-branes to form by quantum fluctuations, their $C_{p\,
+ \, 1}$-flux has to vary by high-dimensional Gauss law at least
within some of the compactified dimensions orthogonal to the
parent branes.  For instance, the variation needed to produce closely
separated daughter branes, say by a distance $M_s^{-1} \, \ll \, L \,
\ll l_{\bot}$, is unlikely to be achieved by de Sitter quantum
fluctuations during brane inflation, since the characteristic
fluctuation wavelength is greater than all the compactified
dimensions, $l_\bot \ll H^{-1}$.

The same result can be easily understood in the language of the
four-dimensional Kaluza-Klein (KK) expansion.  In this language, the
daughter branes act as sources for an infinite tower of massive KK
excitations of the $C_{p\,+\,1}$ form field. $D_{p}$-branes cannot be
formed by quantum fluctuations, unless these fluctuations excite
massive KK states. In order for the RR fluctuations to be sufficient for
the formation of daughter branes with separation $L$, the
gradients of $C_{p\,+\,1}$ must be of the order of
$1/L$. Translated into the KK language, the excited KK must
fluctuate at least by $\sim 1/L$.  However, the mass of the lightest
KK state $m_{KK} \sim 1/l_\bot$ is much bigger that the value of the
Hubble parameter; hence, their fluctuations are exponentially
suppressed, at least as
\begin{equation}
\label{kkfl}
 \sim \, H\, {\rm e}^{-(1/Hl_\bot)}.
\end{equation}
From Eq.~(\ref{scales}), $Hl_\bot\ll 1$, indicating that there will be
essentially no daughter brane formation.
 
The only defects which are not covered by this argument are the branes
that wrap around {\it all} compact dimensions, including the ones
transverse to the parent branes.  However, the formation of such
branes is very costly in energy, since it requires extending the
defect cores into the bulk.  Such a process may occur as a rare
quantum or thermal fluctuation, but it is very strongly
suppressed.\footnote{Spontaneous nucleation of defects by quantum
tunneling in de Sitter space has been discussed in
\cite{Rama}.}

Tachyon variation and RR fields in the bulk can be avoided if the
daughter branes form in close brane-anti-brane pairs, or in the form
of closed loops or surfaces of size comparable to the brane thickness
(which is set by the string scale $M_s$).  However, this would require
variation of $\chi$ on a distance scale $\sim M_s^{-1}$ along the
large dimensions in the parent branes, which would in turn require
expansion rates $H\gtrsim M_s$.  So high an expansion rate is
impossible during the brane inflation [see Eq.~(\ref{scales})].

\section{Solitonic analogies}. 

Solitonic analogies are sometimes useful for understanding the
complicated $D$-brane dynamics. Such analogies have been discussed in
detail in \cite{DVi}. Here we shall illustrate the points we made in
the preceeding Section with an example in which the role of the parent
$D$-branes is played by a domain wall-anti-wall system, the role of
the daughter defects is played by global cosmic strings, and the role
of the bulk RR fields is played by the Goldstone phase.

Formation of low-dimensional daughter defects during the annihilation
of higher dimensional "parents" can occur in systems in which the
order parameter (the Higgs field), responsible for the daughter brane
stability, vanishes in the cores of the parent branes \cite{DV}.  Just
as for $D$-branes, the daughter defects are then the knots of the
tachyon that is condensing during the parent brane annihilation.

In our example, the Higgs field $\Phi$ responsible for the strings
transforms under a global $U(1)$ symmetry, $\Phi \rightarrow
e^{i\alpha}\Phi$.  Outside the domain walls, the Higgs field is in the
vacuum state, $\Phi \, = \, \eta \, e^{i\theta}$, with $\theta={\rm
const}$, while in the wall cores $\Phi=0$.  A cosmic string carries
the topological charge
\begin{equation}
\label{winding}
n\, = \, {1 \over 2\pi} \, \oint\, d\theta, 
\end{equation}
where the integral is taken around a closed loop encircling the
string. Thus, for the formation of a string it is necessary that the
phase of the Higgs field VEV winds non-trivially along the loop.
Formation of such configurations requires that there be a mechanism
which supports zeros in the Higgs VEV at least in some regions of
space, and at the same time allows the phase of the VEV to change
randomly around these zeros.  One example of such a mechanism is
provided by a thermal phase transition with a spontaneous breaking of
a $U(1)$ symmetry (for a review see \cite{AV}).  In this situation,
the symmetry is restored at high temperatures and the Higgs VEV
vanishes.  This happens because the state with a zero Higgs VEV
corresponds to a minimum of the free energy at sufficiently high
temperatures. Hence, zeros in the Higgs VEV are supported by
temperature effects. When the Universe cools below certain critical
point, the minimum turns into a local maximum, the Higgs field becomes
tachyonic and condenses into the true vacuum with a broken symmetry.
During the tachyon condensation, thermal effects force the phase of
the Higgs VEV to randomly fluctuate over distances larger than the
correlation length, and these fluctuations produce topological knots,
the global strings. The typical distance between the strings is set by
the correlation length of the phase fluctuations.

Let us now turn to the formation of strings in our example of
wall-anti-wall annihilation.  We assume the temperature is $T=0$, so
there are no thermal effects, and the Higgs VEV is in the vacuum
everywhere outside the walls.  That is, outside the walls the tachyon
is already in the vacuum state. However, the Higgs field is forced to
zero inside the domain wall cores.  When wall and anti-wall are
brought on top of each other, the tachyon starts condensing. However,
for the tachyon to form topological knots, it is essential that the
phase of the Higgs winds {\it outside} the walls. Thus, if we start
with a configuration of a uniform phase everywhere outside the walls,
we can only expect (at best) to form tiny string loops of size
comparable to the wall width.

To make the analogy with the $D$-brane picture more transparent, let
us rewrite the topological charge of global cosmic strings as the
"electric" charge under a certain two-form field $B_{\mu\nu}$, which
plays the role analogous to that of the RR two-form for $D$-branes.
The two-form field $B_{\mu\nu}$ can be introduced by the following
well-known identification:
\begin{equation}
\epsilon_{\mu\nu\alpha\beta}\partial^{\nu}B^{\alpha\beta} \, = \, 
\partial_{\mu}\theta.
\label{2f}
\end{equation}
The topological charge (\ref{winding}) under $\theta$ now becomes an
electric charge under $B_{\mu\nu}$. As indicated above, in order for
strings to form, it is not enough that the Higgs VEVs on the wall are
uncorrelated, but it is essential that the phase also winds outside
the wall.  In the RR language, this means that $B_{\mu\nu}$ must
assume corresponding ``electric field'' configuration in the bulk.  

Essentially the same remarks apply to daughter brane formation.  We
can think of the bulk space outside the parent branes as having the
tachyon $\chi$ already in its vacuum manifold ${\cal M}$.  Indeed,
regions where $\chi$ develops a VEV in ${\cal M}$ become part of the
bulk.  Since the dimensions transverse to the parent branes are
compact, the variation of $\chi$ in those dimensions is strongly
suppressed during the brane inflation.

\section{Other defect formation mechanisms}

\subsection{Bulk (p)reheating}

We now consider some other defect formation mechanisms in brane
inflation.  The details of brane annihilation process are not well
understood. The colliding brane and antibrane are expected to pass
through one another and oscillate several times before finally
annihilating. The whole process is likely to take time $\tau >>
M_s^{-1}$, and we may well have $\tau\gtrsim l_\bot$. A substantial
part of the energy released in this process will be carried away in
the form of closed string states, that is, gravitatioanal, RR, and
dilaton waves. The resulting RR fluctuations in the bulk may induce
the formation of daughter branes. The cores of these daughter
branes will come from the core region of the annihilating parent
branes.

An alternative version of this scenario is a multi-step brane
annihilation. The colliding stacks of branes and anti-branes do not
have to annihilate all at once. The energy released in the first
annihilation can generate RR fluctuations in the bulk, so that
hedgehog field configurations will appear with their centers lying
within the cores of still existing parent branes. The daughter branes
can then be produced in subsequent annihilations.

To create an infinite network of cosmic strings with a characteristic
length scale $L\lesssim l_\bot$, we need RR fluctuations of wavelength
$\sim L$ and magnitude $dC\sim 1/L$. The corresponding energy density
(in 10 dimensions) is
\beq
\rho^{(L)}_{RR}\sim M_s^{2+d_\parallel +d_\bot}L^{-2}.
\eeq
On the other hand, the energy density obtained if the total energy of 
the annihilating branes is uniformly spread over the transverse dimensions 
is
\beq
\rho^{(tot)}_{RR}\sim M_s^{4+d_\parallel} l_\bot^{-d_\bot},
\label{rhoRR}
\eeq
Since only a fraction of this energy can go into modes of wavelength $L$,
we should have $\rho^{(L)}_{RR}\lesssim\rho^{(tot)}_{RR}$, and thus
\beq
L\gtrsim l_\bot (M_s l_\bot)^{{d_\bot-2}\over{2}}.
\label{Ldbot}
\eeq
For daughter brane formation, we need $L\lesssim l_\bot,$ and it
follows from (\ref{Ldbot}) that this is possible only for $d_\bot\leq
2$. This simple analysis suggests that, for $d_\bot=1$ or 2, daughter
$D$-brane formation is energetically possible, but whether or not it
actually happens depends on the annihilation timescale $\tau$ and on
the power and spectrum of the emitted RR waves. 

The situation may be different in models where the extra
dimensions are strongly warped. For example, in the KKLMMT model of Kachru
{\it et. al.} \cite{string1} the brane annihilation occurs in a deep
gravitational potential well. The energy released in the annihilation
may then be localized in a region $l^{(eff)}_\bot\ll l_\bot$. In the
limit when $l^{(eff)}_\bot\sim M_s^{-1}$, this region may heat up to a
temperature $\sim M_s$ and defect formation may be efficient.

\subsection{Resonant formation of $F$-strings}

A completely different mechanism of defect formation in brane
annihilation has been discussed in \cite{DVi}.  There, it has been
shown that $U(1)$-gauge fields which are massless on the parent branes
get resonantly excited in the process of brane annihilation.  This
mechanism works as follows. When tachyon condenses, it higgses one
combination of the original $U(1)\times U(1)$ group, which we called
$A$. Magnetic flux tubes of this gauge field carry RR charge and no
NS-NS charge and are D-strings.  It was argued in \cite{DVi}, that the
tachyon condensation must at the same time render the orthogonal gauge
field $\tilde{A} \, = \, A^{(1)} \, + \, A^{(2)}$ non-dynamical.  This
is also clear from the fact that, first, this field is not Higgsed,
and second, there must be no open string excitations about the
tachyonic vacuum. So the wave-function of $\tilde{A}$ vanishes as a
result of tachyon condensation.  In a sense, $\tilde{A}$ becomes
infinitely strongly coupled and disappears from the spectrum.  We have
shown that during this process there is an instability towards
generation of an $\tilde{A}$-electric field.  After the branes
dissolve, this gauge field cannot exist in the bulk, so it must be
squeezed into electric (and possibly magnetic) flux tubes.
Electric flux tubes carry NS-NS 2-form $B_{\mu\nu}$ -charge and are
equivalent to the fundamental strings ($F$-strings),

The gauge field amplification has been studied in Ref.\cite{DVi} in a
simple model of an exponential brane-anti-brane decay with a time
constant $\tau$.  It has been shown there that the energy density of
the field (in the parent brane world volume) grows as
\beq
\rho(t)\sim (\tau t)^{-2-(d_\parallel/2)}e^{t/\tau},
\label{rho}
\eeq
and the main contribution to the energy is given by the modes of wavelength
\beq
\l(t)\sim (\tau t)^{1/2}.
\label{lt}
\eeq
The growth is cut off when $\rho(t)\sim M_s^{4+d_\parallel}$, and the
back-reaction becomes important, that is, at $t\sim
(4+d_\parallel)\tau\ln (M_s\tau)$. The characteristic length scale of
field variation at that time is
\beq
l_*\sim \tau [(4+d_\parallel)\ln(M_s \tau)]^{1/2}.
\label{l*}
\eeq

After brane-anti-brane annihilation, we expect a network of
$F$ strings to be formed.  These strings can of course be
infinite or very long in the large dimensions.  The initial scale $L$
of the networks will be set by the characteristic wavelength of the
stochastic electric field, $\l_*$.  We expect the decay
time $\tau$ to be at least a few times greater than $M_s^{-1}$; then
it follows from Eq.~(\ref{l*}) that $l_*$ is greater than $M_s^{-1}$
at least by an order of magnitude.

The character of the $F$-string network produced in this way may
be different from the Brownian networks resulting from the Kibble
mechanism \cite{VV84}. If the typical electric flux through an area
$\sim l_*^2$ is substantially greater than the unit flux trapped in a
string, then the strings will be formed in bundles containing many
strings each. This phenomenon has been observed in simulations of
gauge string formation in the Higgs model where the dominant mechanism
was due to fluctuations of the magnetic field \cite{HR00,SBZ02}.

We note that electric fields pointing into the $d_\parallel$
compactified directions of the parent branes get amplified as
well. This leads to the formation of winding $F$-strings, wrapped
around the compact dimensions. These winding modes correspond to
particles of mass $m\sim M_s^2 l_\parallel$ and are dangerous if
long-lived. They can be avoided if the parent branes are 3-branes with
$d_\parallel=0$, or if the compactified space of the branes has a
trivial first homotopy group, $\pi_1({\cal M}_\parallel) =I$.

The gauge field $A$ orthogonal to ${\tilde A}$ can also be resonantly
excited. The magnetic field of $A$ can then be squeezed into magnetic
flux tubes. These are the $D$-strings. The growth of $A$ at the onset
of brane annihilation is described by the same
Eqs.~(\ref{rho}),(\ref{lt}). However, as the tachyon expectation value
grows, so does the mass of $A$. As a result, the amplification of $A$
is cut off at $t\sim \tau$. The corresponding energy density,
$\rho\sim \tau^{-4-d_\parallel}$, may be insufficiend for the
$D$-string cores, but more energy may be pumped up if the parent
branes oscillate a number of times prior to annihilation, or if the
annihilation is very fast, $\tau\sim M_s^{-1}$. Formation of an
infinite $D$-string network by this mechanism is likely to be
model-dependent.

To summarize, there are two gauge fields, $A$ and $\tilde{A}$, living on
the world volume of the unstable $D$-brane. One of them ($A$) is
Higgsed by the tachyon condensate, and the corresponding magnetic flux
tubes are $D$-strings. The other field $\tilde{A}$ is not Higgsed, but
instead its norm gets diminished by the tachyon. This process creates
electric flux tubes in $\tilde{A}$ which are the fundamental
$F$-strings. Both $D$ and $F$-string networks can be formed as a
result of brane annihilation.  The evolution of these networks will be
discussed in Section VII.

\section{The nature of $D$-strings} 

 In this Section we shall point out some peculiar properties of
 $D$-strings, and show that from the point of view of four-dimensional
 observers they share properties of both local and global strings.
 The role of a cosmic $D$-string in four dimensions can be played by
 any $D_{1\, + \, n}$-brane that is wrapped on $n$-dimensional compact
 manifold and has a one-dimensional projection on the three
 non-compact dimensions.  For instance, it can be simply an unwrapped
 $D_1$ brane; a $D_3$ brane wrapped on two compactified dimensions,
 and so on. The four-dimensional stability of such cosmic $D$-strings
 is due to the four-dimensional two-form projection $C_{\mu\nu}$ of
 the original $C_{2 + n}$ form, where the remaining $n$ indices take
 values in the $n$-dimensional compact space on which the brane wraps.
 Since, according to (\ref{2f}), $C_{\mu\nu}$ in four dimensions is
 equivalent to a Goldstone field, cosmic $D$-strings can be thought of
 as global cosmic strings, of the axionic type, with some crucial
 differences.\footnote{Another connection with axionic type string was
 made in
\cite{dtstrings}, were it was conjectured that, from the point of
view of 4D supergravity theory, $D$-strings are the $D$-term
strings. In this language, the tachyon is the Higgs that compensates
the non-zero $D$-term potential. A detailed analysis of this
conjecture is given in \cite{dtstrings}.} 

To make this connection more
explicit, consider first the field strength and the brane coupling of
the RR form in ten dimensions (we set the string coupling equal to
one),
\begin{equation}
\label{tenc1}
M_s^{8} \, \int_{10}\, (d\, C_{2+n})^2 \, + \, M_s^{2\, + \, n} \,
\int_{2\, + \, n} \, C_{n+2}.
\end{equation}
Reducing this to four dimensions, and keeping only the zero mode $C$, we get 
\begin{equation}
\label{tenc}
M_p^2 \, \int_4\, (d\, C)^2 \, + \, M_s^{2\, + \, n} \,
V_{\parallel}\, \int_2 \, C,
\end{equation}
where we have used the fact that the four-dimensional Planck mass is
given by $M_p^2 \, = \, M_s^8\, V_\parallel V_{\bot}$.  Hence the effective charge
of the comic string with respect to the canonically normalized $C$
field is related to the string tension, 
\begin{equation}
\label{stringtens}
T_s^{(D)} \, \sim \, M_s^{2+n} V_{\parallel} 
\end{equation}
This is in contrast with the ordinary global strings that are formed
by spontaneous breaking of a global $U(1)$-symmetry at some scale
$\eta$. Let $\phi \, = \, \eta \, {\rm e}^{i\theta}$ be the scalar
that forms such a string. Rewriting the Goldstone phase in terms of a
dual two-form field $C_{\mu\nu}$ (we reserve the traditional notation
$B_{\mu\nu}$ for the NS-NS 2-form field) , the analog of the action
(\ref{tenc}) can be written as
\begin{equation}
\label{tenglobal}
\eta^2 \, \int_4\, (d\, C)^2 \, + \, \eta^2\,  \int_{2} \, C.
\end{equation}
From here it follows that the $B$-charge of the global cosmic string
is simply the square root of the string tension, $\sqrt{T_s}\sim
\eta$, without additional $M_p$ suppression.  

Comparison of the two actions, (\ref{tenc}) and (\ref{tenglobal}),
reveals both the similarities and the crucial differences between the
$D$ and ordinary global strings.  Just like for ordinary global
strings, the energy of the three-form field strength in transverse
directions diverges logarithmically,
\begin{equation}
\label{dben}
\int_3 d\, C^2 \, \propto \, {\rm ln R}.
\end{equation}
For $M_s \ll M_P$, however, this divergence is a negligible correction
to the string mass on the present Hubble horizon, $R\sim t_0$, because the
$C_{\mu\nu}$ charge is suppressed by the four-dimensional Planck mass
relative to the string tension,
\beq
{\Delta T_s^{(D)}\over{T_s^{(D)}}}\sim {T_s^{(D)}\over{M_p^2}} 
\ln (M_s t_0) \ll 1.
\eeq

The above is the contribution of the zero mode RR axion to the 
effective four-dimensional string tension. The total contribution from the 
higher KK modes of the RR fields is greater, but only zero mode 
corresponds to a four-dimensional long range field. 

The coupling of the string to the $C_{\mu\nu}$ field is of the same
strength as the one to the four-dimensional zero-mode graviton (and
dilaton).  The relevant ratio that controls the strength of the
interaction is $T_s^{(D)}/M_p^2$, which is small for $T_s^{(D)}\, 
\ll \, M_p^2$.
Because of this very small charge, there are other crucial differences
from the ordinary global strings. The dominant energy loss mechanism
for the conventional global cosmic strings
%, formed by a global $U(1)$ symmetry breaking at scale $\eta$, 
is due to the Goldstone boson radiation \cite{Davis,VV87'}.  According
to (\ref{tenglobal}), the Goldstone coupling to such strings is set by
the symmetry breaking scale (i.e., the tension scale) $\eta$, and for
light strings is very much enhanced relative to the gravitational
coupling.  This is not the case for $D$-strings, for which the
$C_{\mu\nu}$ field couples with the gravitational strength. This is
easy to understand if we recall that both the graviton and the RR
fields (together with the dilaton) come as zero modes of closed
strings, and in the BPS limit their exchanges exactly cancel each
other.  The rate of energy loss by oscillating string loops due to RR
radiation is therefore comparable to that due to gravitational
radiation.  Hence, $D$-strings are sort of hybrid objects. On the one
hand, like global strings (and unlike more conventional gauge
strings), they exert long range RR fields, but on the other hand, the
RR charge is gravitational and does not provide a dominant energy loss
mechanism.

It is well known that in the presence of anomaly ordinary global cosmic
strings become boundaries of domain walls. For cosmic $F$-strings in heterotic
theory this is usually the case\cite{wittencosmic}. 
Such an effect as a possible source for $D$-string instability
was pointed out in \cite{joe}, and was analyzed in \cite{dtstrings}
from the 4D supergravity point of view. 
The bottom-line of the latter analysis is 
the following. As seen from the four-dimensional $N=1$ supergravity theory, 
the $U(1)$-symmetry in question is non-linearly realized on two fields. 
One is the phase of the tachyon $\theta$, and the other is the axion
$a$, which is dual to the (effective) RR two-form field $C$. Under $U(1)$,  
the axion shifts in the following way: 
\beq
a\, \rightarrow \, a\,  + g\, Q_a \alpha, 
\eeq
where $\alpha$ is the gauge transformation parameter, 
$g$ is the gauge coupling, and $Q_a$ is the axionic
charge, given by 
\beq
 Q_a \, = \, {\xi \over M_p^2}. 
\eeq
Here, $\xi$ is of order the $D$-string tension $T$.\footnote{
More precisely, if the $D$-string is a $D_1$-brane, then $\xi = T/2\pi$.}
Since axion is decoupled in the $M_P 
\rightarrow \infty$ limit (an infinite compactification volume), in this limit
the $U(1)$-anomaly must cancel among the massless fermions. 
For finite $M_p$, however, the story is different, since the fermionic
$U(1)$-charges may acquire a shift of order $\xi/M_p^2$, and may contribute
to the chiral anomaly.
In such a case, the consistency of the theory requires that the 
anomaly must be canceled by the 
axionic shift (via Green-Schwarz type mechanism), 
which implies that there is a coupling
\beq
a\, F \wedge F.
\label{affdual} 
\eeq 
Hence, either the fermionic chiral anomaly vanishes on its own,
in which case the coupling (\ref{affdual})is absent, or there is a
fermionic chiral anomaly which is canceled by the axion through the
coupling (\ref{affdual}).  In the former case strings have no axionic
domain walls attached.  In the latter case, the outcome depends on the
presence of mixed chiral fermionic anomalies with other gauge
groups. If such anomalies are present, then the anomaly cancellation
will require couplings of the form (\ref{affdual}) with all the gauge
groups in question. The axion can then get a mass from instantons, and
as a result the $D$-strings will become boundaries of domain walls.
  
Finally, let us point out that $D$-strings can be superconducting, and
thus carry the observable properties of the ordinary superconducting
cosmic strings\cite{wittensuper}.  This is because they may get
connected to a surviving p-brane 
%\footnote{Obviously, the surviving $p$-brane 
%that localizes ordinary particles can have $p \, > \, 3$,  in which case
%it will have some compactified dimensions.}
with open strings whose end points are charged under the gauge group
living on the brane. Massless exitations of these open strings can
carry a current along the $D$-strings.  For charge carriers to be
massless, the $D$-string must be on top of the surviving brane. This
may (usually does) create a potential instability of breakage of the
$D$-string on the surviving brane. 
If $D$-strings are stabilized away from the surviving brane, the
charge carries will be heavy, but they can still be excited by the
cosmic magnetic field. In such a case the signatures are very
different from the superconducting strings with massless charge
carriers\cite{hill}. 
In this case the strings can still break by tunneling into 
monopole-anti-monopole pairs\cite{joe}.
The charge carrier masses in this case will probably be too large for 
any appreciable
currents to develop. The
properties of strings in this case are similar to those of
necklaces (see section VIII.B). 

The tension of the fundamental ($F$) strings is $T_s^{(F)}\sim M_s^2$,
and can be much smaller than that of $D$-strings if $l_\parallel\gg
M_s^{-1}$. The $F$-strings are charged under the NS-NS two-form
$B_{\mu\nu}$, and in this respect are similar to $D$-strings.  An
important property of $F$-strings is that they can end on the
surviving $D$-branes. For instance, an $F$-string intersecting a
$D$-string can break into two, with the ends of the two resulting
strings attached to the $D$-string
\cite{polchinski}.  In the exact supersymmetric limit, for parallel
$D$-brane and $F$-string, this is a dynamically preferred process:
the energy of the system is substantially reduced when the $F$-string
is swallowed by the brane. In the absence of supersymmetry, the situation
may change, because the $D$-string tension gets renormalized, and the
dilaton and RR fields may acquire masses, resulting in a non-trivial
interaction potential between $F$ and $D$-strings. It is conceivable
that as a result the $F$-string breaking process will be suppressed,
so that intersecting $F$ and $D$-strings will simply pass through one
another. (We note, however, that this would require a supersymmetry
breaking scale comparable to the string scale.) Below, we shall
consider both possibilities.

\section{String evolution}

In this Section, we assume that stochastic networks of $D$ and/or
$F$-strings are formed in brane annihilation and study the subsequent
evolution of these networks.
%with different assumptions about the
%nature of the interaction between them.

\subsection{Independent networks}

We shall first consider the evolution of an independent string
network, disregarding the interaction of $D$ and $F$ networks with one
another and with the surviving branes. This may be applicable, for
example, if $F$-string formation is somehow suppressed and $D$-strings
cannot break on the surviving branes. Another possibility is that
$D$-string formation is suppressed and the $F$-network is localized by
bulk gravitational forces in a region with no surviving branes.

 The strings initially form a stochastic network
localized in the plane of the parent branes.  Later they may depart
from that plane, under the action of bulk fields.
Since the extra dimensions are compact, the gravitational effects of
the strings will be the same as those of ordinary cosmic strings.
However, the string evolution can be significantly modified.

A very important role in string evolution is played by reconnections
of intersecting strings.  These reconnections are responsible for
closed loop formation and for the straightening of long strings on the
horizon scale.  Two string segments moving towards one another almost
inevitably intersect in $3D$.  However, it is much easier to avoid
intersection in higher dimensions.  As an illustration, consider two
pointlike particles moving towards one another in a $1D$ universe.  If
the universe is truly $1D$, then the particles inevitably collide.
Now suppose the universe is $(d_\perp +1)$-dimensional, with $d_\perp$
compact dimensions of size $l_\perp$.  Then, if the effective size of
the particle is $\delta\ll l_\perp$, the collision probability is
\beq 
p\sim (\delta/l_\perp)^{d_\perp} \ll 1.
\label{p}
\eeq
Similarly, we expect the $D$-string reconnection probability to be given
by (\ref{p}), with $d_\perp$ being the number of dimensions transverse
to the string world-volume and $\delta\sim M_s^{-1}$.  3-dimensional
observers will not be aware that the strings often miss one another in
extra dimensions.  From their point of view, the strings will
intersect, but fail to reconnect.  The effect of extra dimensions on
string evolution can therefore be simulated by simply introducing a
reconnection probability $p\ll 1$.

For $F$-strings, Eq.~(\ref{p}) is replaced by
\beq 
p\sim (M_s l_\perp)^{-d_\perp}(M_s l_\parallel)^{-d_\parallel}.
\eeq
Hence, the reconnection probabilities for $F$ and $D$-strings are
generally different. Together with the different tensions, this may
result in substantial differences in evolution.

Note that in models with strongly warped bulk geometries, like
the KKLMMT model \cite{string1}, strings may be confined
to the bottom of the potential well, so the effective $l_\bot$ may be
much smaller than the actual size of extra dimensions.

So far, numerical simulations of string evolution in an expanding
universe have been performed
assuming $p=1$.  They showed scale-invariant evolution, with the
characteristic curvature radius of long strings and the typical
inter-string separation both comparable to the horizon, $L\sim t$, and
a large number of small closed loops of sizes $\ll t$.  We expect that
a small reconnection probability will slow down the string evolution,
so that the scale of the network $L$ will be smaller than $t$.  
If $v$ is the characteristic
long string velocity, then the number of encounters of any given long
string with other long strings per Hubble time is $N\sim vt/L$.  In
order for the strings to be able to keep up with the Hubble expansion
and adjust their shape to the increasing size of the horizon, these
encounters have to result in one or few reconnections.  This means
that $Np\sim 1$, and thus \cite{Kibble85}
\beq
L\sim pvt.
\label{pvt}
\eeq
If the motion of strings is relativistic, as in the case of ``usual''
strings in $(3+1)$ dimensions, then $v\sim 1$ and $L\sim pt$.
However, a small reconnection probability may result in accumulation
of small-scale wiggles.  This would increase the effective mass per
unit length of strings and reduce their effective tension.  The string
velocity $v$ will then also be reduced. 

The effect of a small reconnection probability on string evolution has
been studied in flat-space simulations \cite{SV90}. Surprisingly, the
$p$-dependence obtained from the simulations is different from
(\ref{pvt}). A decrease of $p$ from 1 to about 1/3 has little effect
on the string scale $L$. As $p$ is decreased further, $L$ does go
down, but not as fast as Eq.~(\ref{pvt}) suggests. A numerical fit to
the data gives
\beq
L\propto \sqrt{p}
\eeq
in the range $0.05<p<0.3$.
A possible explanation suggested in \cite{SV90} is that due to the
presence of small-scale wiggles, long strings may have many
opportulities to reconnect in each encounter. We expect though that
the scaling (\ref{pvt}) should set in when $p$ gets sufficiently small.
It would be interesting to perform numerical simulations of string
networks with $p\ll 1$ in an expanding universe and to analyze the 
impact of a small $p$ on the observational effects of strings.

Oscillating loops of string will lose their energy by gravitational,
RR, and dilaton radiation. The energy loss rates in all three channels
are comparable to one another, as long as the dilaton and the RR
fields can be regarded as massless. However, the dilaton must have a
nonzero mass $m_d$, and the dilaton radiation is suppressed after the
typical loop size gets larger than $m_d^{-1}$.\footnote{Constraints on
the dilaton mass and string tension resulting from the observational
bounds on dilaton decays have been discussed in \cite{Damour97}. These
bounds may need a revision here, because of the modified string
evolution.}

We note finally that, since the evolving string network is no longer
confined to the initial plane of the branes, string intersections will
occasionally result in formation of small closed loops, winding modes,
wrapped around the compact dimensions.  $3D$ observers will perceive
such loops as point particles of mass $m\sim T_s l_\bot$.  It is
important to estimate the rate of their production, since they can
potentially overclose the universe.  Winding modes do not exist if the
extra dimensional manifold has trivial $\pi_1$.

\subsection{Interacting networks}

Suppose now that both $D$ and $F$-string networks are formed.
$F$-strings will then break as they intersect with $D$-branes
with the break points carrying gauge charges attached to the
branes. Multiple intersections between Brownian $D$ and $F$-strings
will result in chopping up of all $F$-strings into segments with
their ends attached to the $D$-network. If this were the end of the
story, the resulting network would resemble a $Z_3$ string network,
with three strings joined at each vertex \cite{VV87}. However, there
is one further crucial point to consider.

In realistic brane inflation scenarios, some of the original 3-branes
survive annihilation.  In particular, there should be at least one such
brane which we now inhabit. A string segment passing through such a brane
will generally break into two segments, with their ends attached to
the brane.  The resulting $F$-segments will connect $D$-strings to the
branes and will carry $3D$-world gauge charges at the ends connected to the
3-branes. The charges will be pulled back and forth by the string
tension, and the energy of the $F$-segments will be rapidly dissipated
by radiation of gauge quanta. As a result, $D$-strings will be
connected to the 3-branes by short $F$-segments of length $\sim
l_\bot$. The tension in these segments will cause local $D$-string
vibrations, and the energy will further be dissipated into
gravitational, RR, and dilaton waves, until the
$D$-strings become coincident with the 3-branes. In this limit, the
$F$-segments become massless charged particles living on the strings,
and $D$-strings become very similar to ordinary superconducting
strings. Since $D$-strings are now held on top of the 3-brane, their
reconnection probability is $p\sim 1$, as for ordinary strings in $3D$.

In the above discussion we disregarded the force of interaction
between $D$-strings and 3-branes. This interaction indeed vanishes in
the BPS limit, but supersummetry breaking should give rise to a
non-trivial potential. It is possible, in particular, that the
interaction force is attractive at large and repulsive at small
distances, with an equilibrium position at some finite distance from
the brane, $M_s^{-1}\lesssim r\lesssim l_\bot$. The connector
$F$-strings will pull $D$-strings closer to the branes near the points
of connection. In this scenario, the connectors act as massive point
particles, with $m\gtrsim M_s$, from the $3D$ point of view. The
properties of $D$-strings are then similar to those of ``necklaces'',
where massive monopoles and antimonopoles play the role of beads on
the strings \cite{HK85,BV97}.

\section{Conclusions and discussion}

We have examined possible mechanisms of defect formation at the end of
brane inflation.  As brane-anti-brane pairs collide and annihilate,
lower-dimensional $D$-branes can be formed in their place.  We argued
that the crucial requirement for this to happen is that fluctuations
of the tachyon expectation value and of the RR fields should be
excited in the bulk on a length scale smaller than the size of the
transverse dimensions, $l_\perp$.  Quantum de Sitter fluctuations
during inflation have characteristic wavelength $H^{-1}\gg l_\perp$
and fail to satisfy this requirement.

The required RR fluctuations can be produced by the RR waves excited
during the brane annihilation process. Assuming that $l_\bot \gg
M_s^{-1}$, we have shown that this is
energetically possible only if the number of transverse dimensions is
$d_\bot\leq 2$. The dynamics of this process and whether or not it
actually yields an infinite string network remain to be investigated.
In models with a strongly warped bulk, like the model of Kachru {\it
et. al.} \cite{string1}, the effective value of $l_\bot$ may be much
smaller than the actual size of the extra dimensions and can be as
small as $l_\bot^{(eff)}\sim M_s^{-1}$. Then the above bound on
$d_\bot$ does not apply and $D$-string formation can be very
efficient.

We have pointed out that cosmologically interesting defect networks
can be produced by the mechanism suggested in Ref.~\cite{DVi}.  The
crucial point is the following.  Out of the two massless gauge fields
($A,\tilde{A}$) living in the world volume theory of an unstable
brane-anti-brane pair, one ($A$) is Higgsed by the tachyon condensate.
The corresponding magnetic flux tubes are $D$-strings. The other field
($\tilde{A}$) is rendered non-dynamical by the tachyon VEV.  During
this process, ${\tilde A}$-electric field gets resonantly excited.
This field is then squeezed into electric flux tubes, which are the
$F$-strings. The resonant excitation of the magnetic component of the
field $A$ is less efficient, but can be enhanced if brane annihilation
is preceeded by the collision and multiple oscillations of the parent
branes.  Quantitative conditions for the formation of a $D$-string
network require further study.

We have discussed the physical properties of $D$-strings and pointed
out some important differences from the ordinary cosmic strings. The
tension of $D$-strings, $T_s^{(D)}\sim M_s^2 (M_s
l_\parallel)^{d_\parallel}$ can be much greater than that of
$F$-strings, $T_s^{(F)}\sim M_s^2$, if the compactified dimensions of
the parent branes are large compared to the string scale,
$l_\parallel\gg M_s^{-1}$.  Like ordinary (gauge) cosmic strings,
$D$-strings are vortices of the tachyon carrying a unit magnetic flux
of $A$. At the same time, they carry a long-range RR field $C$ and are
in this respect similar to global strings. An important
difference is that the coupling to $C$ is suppressed by the ratio
$T_s^{(D)}/M_p^2$ compared to the ordinary global strings.  As a
result, RR radiation power from an oscillating D-string loop is
comparable to that of gravitational radiation. This is in contrast to
global strings, for which the Goldstone boson radiation is the
dominant energy loss mechanism.

Once the string networks are formed, their subsequent evolution
crucially depends on the nature of interaction of $D$ and $F$ strings
with one another and with the surviving 3-branes. If both types of
networks are formed, then an $F$-string intersecting a $D$-string can
break into two, with the ends of the two resulting strings attached to
the $D$-string. The same kind of process occurs when $F$-strings
collide with one of the surviving branes. As a result of these
interactions, the $F$-network will be chopped into small segments
connecting $D$-strings to the 3-branes. The energy of the segments
will be rapidly dissipated, so the $D$-strings will become coincident
with the 3-branes, and the segments themselves will turn into massless
charged particles living on the strings. $D$-strings in this scenario
will be very similar to ordinary superconducting strings. If there is
an interaction potential that holds $D$-strings at a finite distance
from 3-branes, the $F$-segments will appear as massive point particles
of mass $\gtrsim M_s$ attached to the strings, as in ``necklaces'',
where massive monopoles play the role of beads on strings.

We have also considered the evolution of a single string network,
assuming that its interaction with the surviving branes is negligible.
The strings then propagate in the
higher-dimensional bulk and can avoid intersections much more easily
than strings in 3 dimensions.\footnote{We should stress, however, 
that in a concrete model, whatever
dynamics will prevent strings from breaking on the surviving brane,
probably will also restrict their motion in full high-dimensional space. 
So we should be careful with applying the above argument.} For 
macroscopic observers, string
intersections will appear as frequent as usual, but on most occasions
the intersecting strings will fail to reconnect.  Thus, the effect of
extra dimensions can be accounted for by introducing a small
reconnection probability, $p\ll 1$.  This is likely to result in the
increased wiggliness of strings and in a smaller inter-string
separation.  
%This properties, however, are subject to supersymmetry
%breaking effects, which may produce non-trivial interaction potential
%in extra dimensions.

The string networks produced in brane annihilation are potentially
observable through gravitational lensing \cite{AV81}, linear
discontinuities on the microwave sky \cite{Stebbins}, gravitational
wave bursts \cite{Damour}, or a stochastic gravitational wave
background \cite{VV}. If $D$-strings are superconducting, the
gravitational wave bursts they produce may be accompanied by gamma-ray
bursts \cite{BHV}. On the other hand, if $D$-strings behave as
necklaces, the annihilations of heavy ``beads'' on the strings may
produce ultrahigh-energy cosmic rays \cite{BV}.

We finally note an intriguing recent observation of two nearly identical
galaxies at redshift $z=0.46$ with an angular separation of 1.9 arc
seconds \cite{string}.  The most plausible interpretation appears to
be lensing by a cosmic string with $T_s/M_p^2 \sim 3.7\times 10^{-7}$
\cite{string}.  This estimate assumes a slowly moving string
orthogonal to the line of sight at a relatively low redshift
$(z\lesssim 0.1)$.  Increasing the string redshift or changing its
orientation leads to a higher estimate for $\mu$, which may be in
conflict with the microwave observation data.  (The most recent
constraint from string simulations is \cite{Shellard} $T_s/M_p^2\lesssim
7\times 10^{-7}$.)  On the other hand, the estimate for $T_s$ can be
decreased due to (relativistic) motion \cite{AV86} or wiggliness
\cite{VV2} of the string.

\medskip
\section*{Acknowledgments}

While this work was in progress, we learned that N. Jones, H. Stoica
and S.-H. Tye have independently reached the conclusion that string
reconnections may be suppressed in braneworld cosmology, resulting in
a denser string network \cite{JSTnew}. The string evolution with a low
reconnections probability was then discussed in \cite{PTW}.

We are grateful to Thibault Damour, Gregory Gabadadze, Shamit Kachru,
Renata Kallosh, Andrei Linde, Rob Myers, Joe Polchinski and
Henry Tye for useful discussions.
The research of G.D. is suported in part by a David and Lucile Packard
Foundation Fellowship for Science and Engineering, and by NSF grant
PHY-0070787.  The work of A.V. was supported in part by the National
Science Foundation.

\end{document}